\documentclass[conference]{IEEEtran}
\IEEEoverridecommandlockouts
\usepackage{cite}
\usepackage{amsmath,amssymb,amsfonts}
\usepackage{algorithmic}
\usepackage{graphicx}
\usepackage{textcomp}
\usepackage{xcolor}
\usepackage{cite}
\usepackage{algorithmic}
\usepackage{algorithm}
\usepackage{enumerate}
\usepackage{subcaption}
\def\BibTeX{{\rm B\kern-.05em{\sc i\kern-.025em b}\kern-.08em
    T\kern-.1667em\lower.7ex\hbox{E}\kern-.125emX}}
\begin{document}

\title{Reconfigurable Intelligent Surfaces-Assisted
Integrated Access and Backhaul\\
{\footnotesize \textsuperscript{}}
%\thanks{Identify applicable funding agency here. If none, delete this.}
}
\author{
 \IEEEauthorblockN{Charitha Madapatha\IEEEauthorrefmark{1}, Behrooz Makki\IEEEauthorrefmark{2}, Hao Guo\IEEEauthorrefmark{1}\IEEEauthorrefmark{3}, Tommy Svensson\IEEEauthorrefmark{1}} 
\IEEEauthorblockA{\IEEEauthorrefmark{1}Department of Electrical Engineering, Chalmers University of Technology, Gothenburg, Sweden\\
\IEEEauthorrefmark{2}Ericsson Research, Ericsson AB, Gothenburg, Sweden\\
\IEEEauthorrefmark{3}Tandon School of Engineering, New York University, Brooklyn, NY, USA\\
charitha@chalmers.se, behrooz.makki@ericsson.com, hao.guo@chalmers.se, tommy.svensson@chalmers.se
}
}
\maketitle

\begin{abstract}
In this paper, we study the impact of reconfigurable intelligent surfaces (RISs) on the coverage extension of integrated access and backhaul (IAB) networks. Particularly, using a finite stochastic geometry model, with random distributions of user equipments (UEs) in a finite region, and planned hierachical architecture for IAB, we study the service coverage probability defined as the probability of the event that the UEs’ minimum rate requirements are satisfied. We present comparisons between different cases including IAB-only, IAB assisted with RIS for backhaul as well as IAB assisted by network controlled repeaters (NCRs).
Our investigations focus on wide-area IAB assisted with RIS through the lens of different design architectures and deployments, revealing both conflicts and synergies for minimizing the effect of tree foliage over seasonal changes. Our simulation results reveal both opportunities and challenges towards the implementation of RIS in IAB. 
\end{abstract}


\section{Introduction}
The rapid growth in data traffic and the advent of technologies like the Internet of Things (IoT) and 5G networks necessitate innovative solutions for wireless communication systems \cite{eref2}. Reconfigurable intelligent surface (RIS) and integrated access and backhaul (IAB) networks emerge as candidate techniques, offering enhanced coverage, capacity, and reliability leading to a new era in wireless network architecture.

RIS is a planar meshed surface with multiple passive reflective surfaces attached to each grid, which can coordinately align its phases using a controller and perform beamforming. Thus, similar to repeaters, RIS will also require a control link. RISs have the potential to be deployed densely in a network with low energy consumption \cite{b1,parambath2024integrating}. Also, the passive nature of the elements in RIS allows them to be coated to any surface without hassle, making its deployment faster \cite{wang2024wideband}. Aligning the phase of RIS towards a particular user equipment (UE) requires channel state information (CSI)  which is regarded as a challenge in highly dynamic propagation scenarios. The programmability of RIS also facilitates energy-efficient network operations by allowing precise control over the radio propagation environment \cite{b2}. This programmability, coupled with its adaptability, makes RIS an interesting enabler for future wireless networks.

IAB is a decode-and-forward relaying method standardized in 3GPP Releases 16-18 where an IAB node provides intermediate connections between, e.g., the IAB donor and the child nodes (in general, IAB can be a multi-hop network but practically it may be limited to 2 or 3 hops due to its latency \cite{fang2021joint, 9548327, madapatha2023constrained}). According to 3GPP specifications, IAB nodes are classified into two categories: wide-area and local-area IAB. These categories differ primarily in terms of their functional capabilities and the extent of network planning required for their deployment \cite{teyeb2019integrated, madapatha2023constrained}. Particularly, compared to local-area IAB, wide-area IAB is a more powerful/capable node used for coverage extension in a wide area.

In particular, RIS may be helpful by providing alternative paths when the direct link from the IAB-donor to IAB-child faces deep fading and/or blockage and makes it easier to avoid such trouble. The dynamic nature of RIS deployments allows for flexibility in addressing environmental challenges such as foliage, rain, and propagation losses. Recent advancements highlight the growing importance of RISs and IAB networks in next-generation communication systems. For instance, RISs have demonstrated significant potential in improving signal strength and energy efficiency through passive beamforming. Integration of RIS with IAB networks capitalizes on these strengths, enabling seamless backhaul and access link operations even under challenging propagation conditions \cite{b1}. Furthermore, the integration of RIS with IAB networks enables dynamic resource management and can improve the overall network energy efficiency. As explored in the literature, the programmability and adaptability of RIS enhance the flexibility of IAB nodes, making them resilient to environmental factors such as rain and foliage. These features allow RIS to provide alternative paths for signal propagation, ensuring reliable connectivity in scenarios where direct links face challenges \cite{b2}. \textcolor{black}{The combination of RIS and IAB also aligns with the objectives of future 6G networks, which aim to optimize resource utilization and energy efficiency \cite{b3}}.

\textcolor{black}{
In this paper, we study the performance of RIS-assisted IAB networks in the cases with different environmental effects. By dynamically managing RIS elements and balancing the load between backhaul and access links, this approach addresses the growing demand for scalable, energy-efficient wireless networks capable of supporting high user densities and diverse application scenarios. This work builds on these advancements, progressing towards a framework for deploying RIS-assisted IAB networks in real-world environments. These advancements reflect the ongoing evolution in network design, particularly with the introduction of RIS-assisted systems. They aim to achieve highly flexible, sustainable, and energy-efficient wireless communication infrastructures, aligning with the broader goals of future 6G networks. This study builds on these developments to explore practical implementations and optimizations for RIS-assisted IAB deployments.}

In particular, we study the service coverage probability as the performance metric. We consider a wide-area setup and consider RIS to improve the coverage area of the IAB in the cases with, e.g., seasonal tree foliage. Moreover, we compare the performance of RIS-assisted IAB networks with cases having only IABs or the cases where network-controlled repeaters (NCRs) are used to assist the IAB network. Also, we investigate the effect of rain loss on the performance of these networks. Our simulations show that the use of RISs/NCRs help to increase the service coverage probabilty in the presence of tree foliage.

% The structure of the rest of the paper is as follows. In section
% II, we discuss the integration of RIS and IAB including user association, modeling of the RIS, channel model. In section III, a comparative study is performed between RIS assited IAB for various deployment strategies in the presence of tree foliage, rain over seasonal variations Finally, in section IV, conclusions are
% drawn based on the study and our future works are discussed.
\section{The Integration of RIS and IAB}

\begin{figure*}
    \centering
\includegraphics[width=1.5\columnwidth]{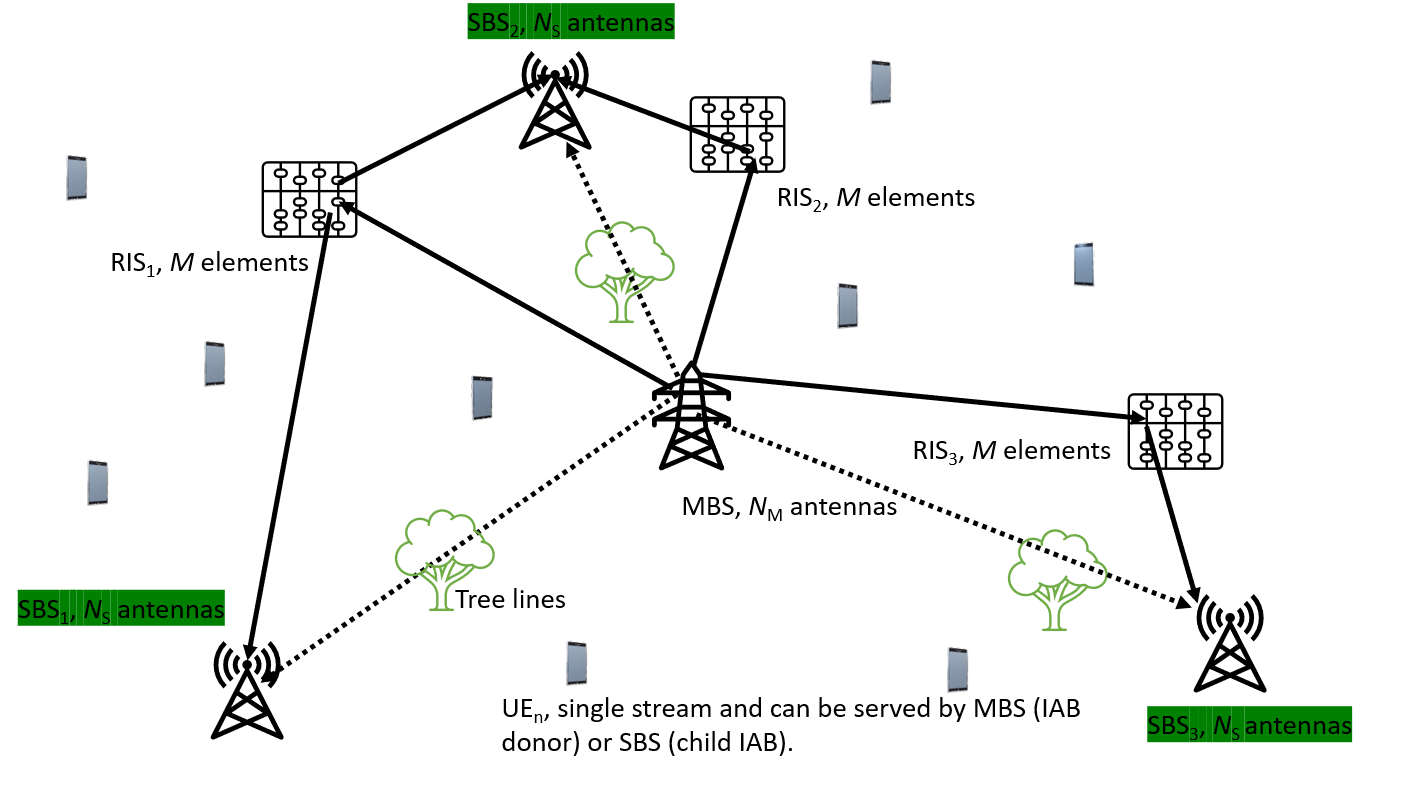}
\vspace{-0.3cm}
    \caption{Illustration of the wide-area IAB assisted with RIS for backhaul links amidst seasonal tree foliage changes.}
    \label{fig:CLOSE_6G}
\end{figure*}
Consider an in-band wide-area IAB network with two-hops in downlink where the IAB donor and IAB child nodes can serve multiple UEs as shown in Fig. \ref{fig:CLOSE_6G}. In contrast to a local-area IAB network, this topology is designed with the intention to extend coverage focusing on sub-urban areas. Here, UEs are served either by the IAB donor or IAB child nodes, while the backhaul link can be either directly from IAB donor to IAB child node or be via RIS depending on the seasonal tree foliage level in the direct backhaul link. In particular, the direct link can be blocked by the dense tree lines while the path via the RIS does not have obstructions from tree lines. This is motivated by the fact that the effect of the tree foliage may change considerably in different seasons where, for instance, the foliage loss may increase signficantly during the spring and summer. 

\subsection{Backhaul via Direct links}
Following the state-of-the-art mmWave channel model given in \cite{ref1, erefour}, the received power at each node can be expressed as
\begin{equation}
    P_{\text{rx}}=P_{\text{tx}}h_{\text{link}}G_{\text{ant}}
    \eta_{\text{path}}\phi_{\text{loss}}.
    \label{eq_a1}
\end{equation}

Here, $P_{\text{tx}}$, $h_{\text{link}}$, $G_{\text{ant}}$ denote the transmit power, Rayleigh fading, and antenna gain, respectively while $\phi_{\text{loss}}$ is the tree foliage loss. Furthermore, path loss denoted by $\eta_{\text{path}}$, in dB, is characterized according to the 5GCM UMa close-in model described in \cite{ref3} as below.

\begin{equation}
\eta_{\text{path}}=32.4+10\log_{10}(d)^{\alpha}+20\log_{10}(f_{c}),
\end{equation}
where $f_c$ is the carrier frequency, $d$ is the propagation distance between the nodes, and $\alpha$ is the path loss exponent. 
Depending on the blockage, line-of-sight (LoS) and NLoS (N: Non) links are affected by different path loss exponents. 

5G and beyond systems are equipped with large antenna arrays which are used to minimize the propagation loss. We use the sectored-pattern antenna array model to characterize the beam pattern and antenna gain, which is given by

\begin{equation}G_{\text{ant}}(\theta) = \begin{cases} G_{\text{main}}&\frac{-\theta_{\text{HPBW}}}{2}\leq\theta\leq\frac{\theta_{\text{HPBW}}}{2} \\
 g_{\text{side}}(\theta)&\text{otherwise.} \end{cases}
\end{equation}
Here, $\theta$ represents the angle between the transmit and receive antennas. Furthermore, $\theta_{\text{HPBW}}$ is the half power beamwidth, and $G_{\text{main}}$ denotes the main lobe gain of the antenna while $g_{\text{side}}(\theta)$ is the side lobe gain \cite{ref1}. Similar to \cite{eref29,eref19,erefour}, we assume an omni-directional beam pattern with UE antenna gain of 0 dB.

We assume that the inter-UE interference is negligible in the downlink with the assumption of sufficient isolation. Particularly, the interference model focuses on the aggregated interference on the access links, caused by the neighboring interferers, which for UE $u$ is expressed as    
\begin{equation} I_{u}= \sum \limits _{ {i,u}\in \Lambda _{i,u}\setminus \{{w}_{u}\}}{P_{i}}  h_{i,u} G_{i,u}{\eta}_{(1\text{m})} \eta_{{x_i,x_u}}\|{\mathbf{x_i}-x_u}\|^{-1},
\label{inter_eq}
\end{equation}
where $i$ denotes the set of BSs excluding the associated BS $w_u$ of user $u$. Also, for SBS $s$, the aggregated interference on the backhaul links is given by
\begin{align} I_{s}= \sum \limits _{ {j,s}\in \Lambda_{j,s}\setminus \{{w}_{s}\}}{P_{j}}  h_{j,s} G_{j,s}{\eta}_{(1\text{m})} \eta_{{x_j,x_s}}\|{\mathbf{x_j}-x_s}\|^{-1},
\label{intersbs_eq}
\end{align}

where $j$ denotes the set of transmitting BSs excluding the associated BS $w_s$ of SBS $s$.

We use a finite homogeneous Poisson point process (FHPPP) denoted by $\Lambda_{T}$ with density $\lambda_{T}$ to model the spatial distribution of the tree lines of length $r$ \cite{gensys6}. The tree foliage loss is estimated using the Fitted International Telecommunication Union-Radio (FITU-R) tree foliage model \cite[Chapter 7]{gensys7}. The model is well known for its applicability in cases with non-uniform vegetation and frequency dependency within the 10-40 GHz range. Particularly, considering both in-leaf and out-of-leaf vegetation states, the tree foliage loss in \eqref{eq_a1} is expressed as  
\begin{equation}
   \phi_{\text{loss}}=\begin{cases}0.39f_{\text{c}}^{0.39}r^{0.25},\;\text{in-leaf}\\0.37f_{\text{c}}^{0.18}r^{0.59},\;\text{out-of-leaf,}\;\end{cases}
   \label{eq_veg_loss}
\end{equation}
where $r$ is the vegetation depth measured in meters.

In our setup, each UE has the ability to be connected to either an MBS or an SBS depending on the maximum average received power. Let $a_{u}\in \{0,1\}$ be a binary variable indicating the association with 1, while 0 representing the opposite. Thus, for the access links 
\begin{align} {a_u} = \begin{cases} 1, & \text {if}~P_{i} G_{z,x}h_{z,x}{\eta}_{(1m)} \eta_{z,x}(\|{\mathbf{z}}-{\mathbf{x}}\|)^{-1} \\
&\qquad \quad ~~\geq P_{j}G_{j}h_{z,y}{\eta}_{(1m)} \eta_{z,y}(\|{\mathbf{z}}-{\mathbf{y}}\|)^{-1}, \\
&\qquad \qquad \quad \forall ~{\mathbf{y}}\in \Lambda _{j}, j\in \{{\mathrm{ m}},{\mathrm{ s}}\}|{\mathbf{x}}\in \Lambda _{i}, \\ 0,& \text {otherwise,} \end{cases}\end{align} \label{assoc_eq1}
where ${i}$, $j$ denote the BS indices, i.e., MBS or SBS. As in (7) for each UE $u$, the association binary variable $a_u$ becomes 1 for the cell giving the maximum received power at the UE, while for all other cells it is 0, as the UE can only be connected to one IAB node. 

Since the MBSs and the SBSs have large antenna arrays and can beamform towards the desired direction, the antenna gain over the backhaul links can be assumed to be the same, and backhaul link association can be well determined based on the minimum path loss rule, i.e., by
\begin{align}
{a_{\text{link}}} = \begin{cases} 1, & \text {if}~\eta_{{\rm {link}}}(\|{\mathbf{u}}-{\mathbf{v}}\|)^{-1}\!\geq \! \eta_{{\rm {link}}}(\|{\mathbf{u}}-{\mathbf{w}}\|)^{-1}, \\
 &\qquad \qquad \qquad \qquad \qquad \forall ~{\mathbf{w}}\in \Lambda_{\text{nodes}}|{\mathbf{v}}\in \Lambda_{\text{nodes}}, \\
 0,& \text {otherwise.} \end{cases} \end{align}

For resource allocation, on the other hand, the mmWave spectrum available is partitioned into the access and backhaul links such that
\begin{equation}\begin{aligned}
    \begin{cases}
    B_{\text{backhaul}}=\psi B,\\
    B_{\text{access}}=(1-\psi)B.
\end{cases}
\end{aligned}
  \label{eq:11}  
\end{equation}

In practice, along with the MBSs which are non-IAB backhaul-connected, a portion of the SBSs may have dedicated non-IAB backhaul connections, resulting in a hybrid IAB network. Therefore, in our deployment, some of the SBSs are IAB backhauled wirelessly and the others are connected to dedicated non-IAB backhaul links. 

Let us initially concentrate on the IAB-type backhauled SBSs. Also, let, $B_{\text{backhaul}}$ and $B_{\text{access}}$ denote the backhaul and the access bandwidths, respectively, while total bandwidth is $B=B_{\text{backhaul}}+B_{\text{access}}$. The bandwidth allocated for each IAB-type wirelessly backhauled SBS, namely, child IAB, by the MBS, i.e., IAB donor, is proportional to its load and the number of UEs in the access link. The resource allocation is determined based on the instantaneous load where each IAB-type backhauled SBS informs its current load to the associated MBS each time. Thus, the backhaul-related bandwidth for the $k$-th IAB node, if it does not have dedicated non-IAB backhaul connection, is given by  
\begin{equation}
{B_{\text{backhaul},k}}=\frac{\psi BN_k}{{\displaystyle\sum_{\forall\;k}}N_k},{     \forall k},
\end{equation}
where $N_k$ denotes the number of UEs connected to the $k$-th IAB-type backhauled node and $\psi\in [0,1]$ is the fraction of the bandwidth resources on backhauling. Therefore, the bandwidth allocated to the $k$-th IAB-type backhauled node is proportional to the ratio between its load, and the total load of its connected IAB donor. Meanwhile, the access spectrum is equally shared among the connected UEs at the IAB node according to
\begin{equation}
{B_{\text{access},u}}=\frac{(1-\psi)B}{{\displaystyle\sum_{\forall\;u}}N_{k,u}}, \forall u,
\end{equation}
where $u$ denotes the UEs indices, and $k$ represents each IAB-type backhauled node. Moreover, $N_{k,u}$ is the number of UEs connected to the $k$-th IAB-type backhauled node to which UE $u$ is connected. Finally, the signal-to-interference-plus-noise ratio (SINR) is obtained in accordance with \eqref{inter_eq} by 
\begin{equation}
\text{SINR}=P_{\text{rx}}/(I_{u}+\sigma^2),
\end{equation}
where $\sigma^2$ is the noise power.

With our setup, the network may have three forms of access connections, i.e., MBS-UE, IAB-type, and non-IAB backhauled SBS-UE. The individual data rates will
behave according to the form in which the UE’s connection has
been established. Particularly, the rates experienced by the UEs in access links that are connected to MBSs or to the IAB type-backhauled SBSs are given by 
\begin{equation}
 {R_{u}} =   \begin{cases}\frac{(1-\psi)B}{N_{m}}\log(1+\text{SINR}(v_{u})), ~\text { if }{w}_u\in \Lambda _{\text{MBS}},\\ \min \bigg (\frac{(1-\psi)BN}{{\displaystyle\sum_{\forall\;u}}N_{k,u}}\log(1+\text{SINR}(v_{u})), \\
 \qquad  \frac{\psi BN}{{\displaystyle\sum_{\forall\;k}}N_k}\log(1+\text{SINR}(v_{b}))\bigg ), \text {if }{w}_u\in \Lambda _{\text{SBS}}, \end{cases}
 \label{eq:14}
\end{equation}
where $k$ represents each IAB-type backhauled SBS connected to the MBS. Then, $m$ gives the associated MBS, $s$ denotes the SBS, and $u$ represents the UEs' indices. 

Depending on the associated cell, there are two possible cases for the data rate of the UEs. First is the case when the UEs are connected to the MBSs, i.e., IAB donor, as denoted by $w_{u} \in \Lambda _{\text{MBS}}$ in \eqref{eq:14}. Since the MBSs have non-IAB backhaul connection, the rate will only depend on the access bandwidth available at the UE. In the second case, the UEs are connected to the IAB-type backhauled SBSs, as denoted by $w_{u} \in \Lambda _{\text{SBS}}$ in \eqref{eq:14}. Here, the SBSs have shared backhaul bandwidth from the IAB-Donor-nodes  i.e., MBSs, and thus the UEs data rates depend on the backhaul rate of the connected IAB-type backhauled SBS as well. Thus, in this case the UE is bounded to get the minimum between backhaul and access rate. 

\subsection{Backhaul via RIS}
Consider an $M$-element RIS assisting the backhaul between IAB donor and IAB child, where the receive signal at the RIS can be expressed in \cite{guo2022comparison} as
\begin{equation}
y = \sqrt{P_{\text{tx}}}\mathbf{g}_c \mathbf{w}_c x + z,
\end{equation}
where $P_{\text{tx}}$ is the transmit power at the donor IAB, $x$ is the transmitted signal with a unit power, $\mathbf{w}_c \in \mathbb{C}^{N_c \times 1}$ is the beamformer of the BS-RIS link and has unit power as an upper bound, and $z$ is the additive Gaussian noise at the receiver side. 

Also, ignoring the direct IAB-donor to IAB-child link as well as the nonlinearities between adjacent reflectors, the equivalent channel $\mathbf{g} \in \mathbb{C}^{1 \times N_r}$ in the RIS-assisted link (BS-RIS-UE) is given by
\begin{equation}
\mathbf{g}_c = \mathbf{g}_{ru} \boldsymbol{\Omega} \mathbf{g}_{br}.
\end{equation}
Here, $\mathbf{g}_{br} \in \mathbb{C}^{M \times N_r}$ and $\mathbf{g}_{ru} \in \mathbb{C}^{1 \times M}$ are the channel between IAB donor-RIS and RIS-IAB child, respectively. Moreover,
\begin{equation}
\boldsymbol{\Omega} = \text{diag}(e^{j\omega_1}, \ldots, e^{j\omega_M})
\end{equation}
is the reflection coefficient matrix of the RIS. With joint BS active and RIS passive beamforming, the backhaul data rate at child IAB node with RIS assistance can be expressed as
\begin{equation}
R = \log\left( 1 + \frac{P \left| \mathbf{g}_{ru} \boldsymbol{\Omega} \mathbf{g}_{br} \mathbf{w}_c \right|^2}{\sigma^2} \right).
\end{equation}

To optimize the RIS matrix $\boldsymbol{\Omega}$, the phase shifts $\omega_M$
are adjusted to maximize the effective channel gain, $\left| \mathbf{g}_{ru} \boldsymbol{\Omega} \mathbf{g}_{br} \mathbf{w}_c \right|^2$ ensuring constructive signal combination at the IAB child.

\section{Simulation Results and Discussion}

This section evaluates the effect of RIS-assistance in mitigating the tree foliage and rain loss on the service coverage probability, $\rho$ of IAB networks. Here, $\rho$ can be expressed using \eqref{eq:14} as $\rho= \Pr(R_\text{u} \ge \beta)$ where $\beta$ represents the data rate threshold. The simulation results and discussions are divided into five parts to provide a comprehensive understanding of the network performance.

\begin{figure}
\centerline{\includegraphics[width=3.4in]{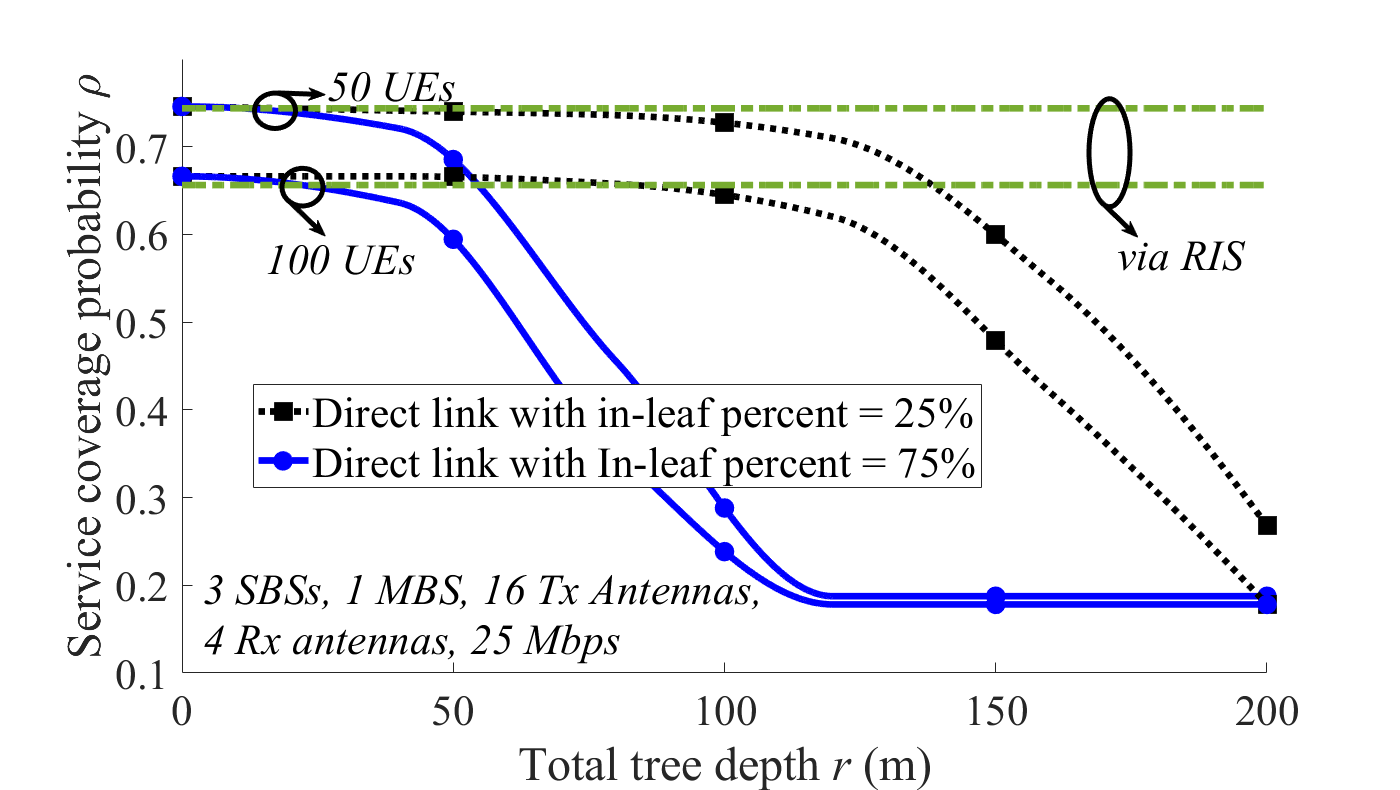}}
\caption{Network service coverage probability as a function of total tree depth.}
\label{stp_beamform}
\end{figure}

\subsubsection{Impact of Tree Foliage on Service Coverage}
Figure 2 illustrates the network service coverage probability as a function of total tree depth, \(r\), for different percentages of in-leaf conditions and the inclusion of RIS-assisted links. The parameters are set to 1 MBS, 3 SBSs, 16 Tx antennas and 4 Rx antennas, and in all cases the minimum UE data rate threshold requirement is kept at 25 Mbps.  For both 50 and 100 UEs, it is observed that the direct link's performance deteriorates significantly as the tree depth increases, especially under 75\% in-leaf conditions. For instance, with a total tree depth of 150 m, service coverage probability drops below 0.3 for the 75\% in-leaf scenario, compared to approximately 0.6 for the 25\% in-leaf case. Also, for small values of total tree depth $r$, the traditional IAB network is depicting similar performance to an RIS-aided network. However, as the $r$ becomes larger, the RIS-aided network outperforms the IAB only backhauled network. This is because, with larger $r$ it is more likely for the backhaul links to be affected by tree foliage than the double path-loss of RIS-aided backhaul link. 

The RIS-assisted links provide substantial improvement in coverage by bypassing the obstructions caused by dense foliage. As seen, service coverage probability for RIS-assisted communication remains above 0.7, even at a tree depth of 200 m, highlighting the robustness of this solution. This improvement underscores the RIS's capability to mitigate coverage issues in foliage-heavy environments.

\subsubsection{Carrier Frequency and User Density Effects}
Figure 3 evaluates the service coverage probability as a function of the number of UEs for two carrier frequencies, 28 GHz and 38 GHz. Across all scenarios, service coverage probability decreases with an increasing number of UEs due to higher competition for resources. However, the performance at 28 GHz consistently outperforms 38 GHz, attributed to lower path loss at lower carrier frequencies. In reality, networks often use multiple frequency bands for complementary purposes (e.g., combining the strengths of 28 GHz for extended coverage with the higher capacity of 38 GHz in small cells). 

As we see, with 60 UEs and a tree line depth of 25 m, service coverage probability is approximately 0.6 for 28 GHz but drops to about 0.3 for 38 GHz. This disparity becomes more pronounced as the user density increases. Moreover, increasing the tree line depth to 75 m significantly affects coverage, reducing service coverage probability by nearly 50\% for both frequencies. These results emphasize the trade-offs between carrier frequency and user density in designing robust IAB networks.

\subsubsection{Rain Intensity and Tree Foliage Combined Effect}
In Fig. 4, we investigate the combined impact of rain intensity and tree foliage on service coverage probability. As rain intensity increases, service coverage probability declines for all configurations, with direct links under 75\% in-leaf conditions showing the worst performance. At a rain intensity of 10 mm/hr, service coverage probability is below 0.2 for this scenario, indicating severe degradation. The inclusion of RIS or SBS mitigates this issue significantly. For instance, RIS-assisted communication with 200 reflecting elements maintains service coverage probability above 0.6 across all rain intensities, demonstrating its resilience. Similarly, SBSs provide consistent improvement, although their performance is slightly less robust compared to RIS. These results suggest that deploying RIS in combination with SBSs can effectively counteract adverse environmental conditions, ensuring reliable service coverage.

\begin{figure}
\centerline{\includegraphics[width=3.4in]{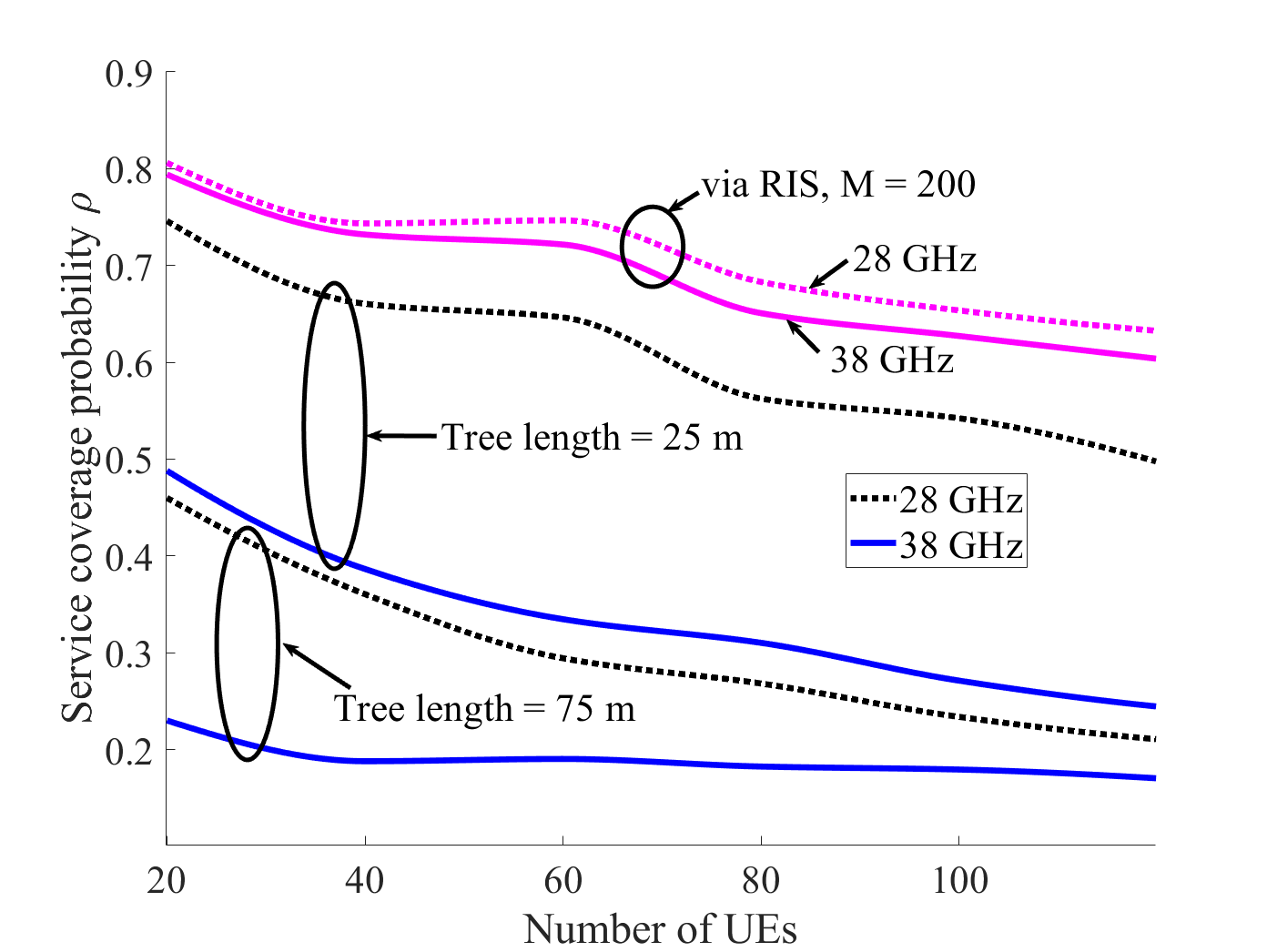}}
\caption{Network service coverage probability for different carrier frequencies as a function of number of UEs,  $N$. }
\label{stp_beamform1}
\end{figure}

\begin{figure}
\centerline{\includegraphics[width=3.4in]{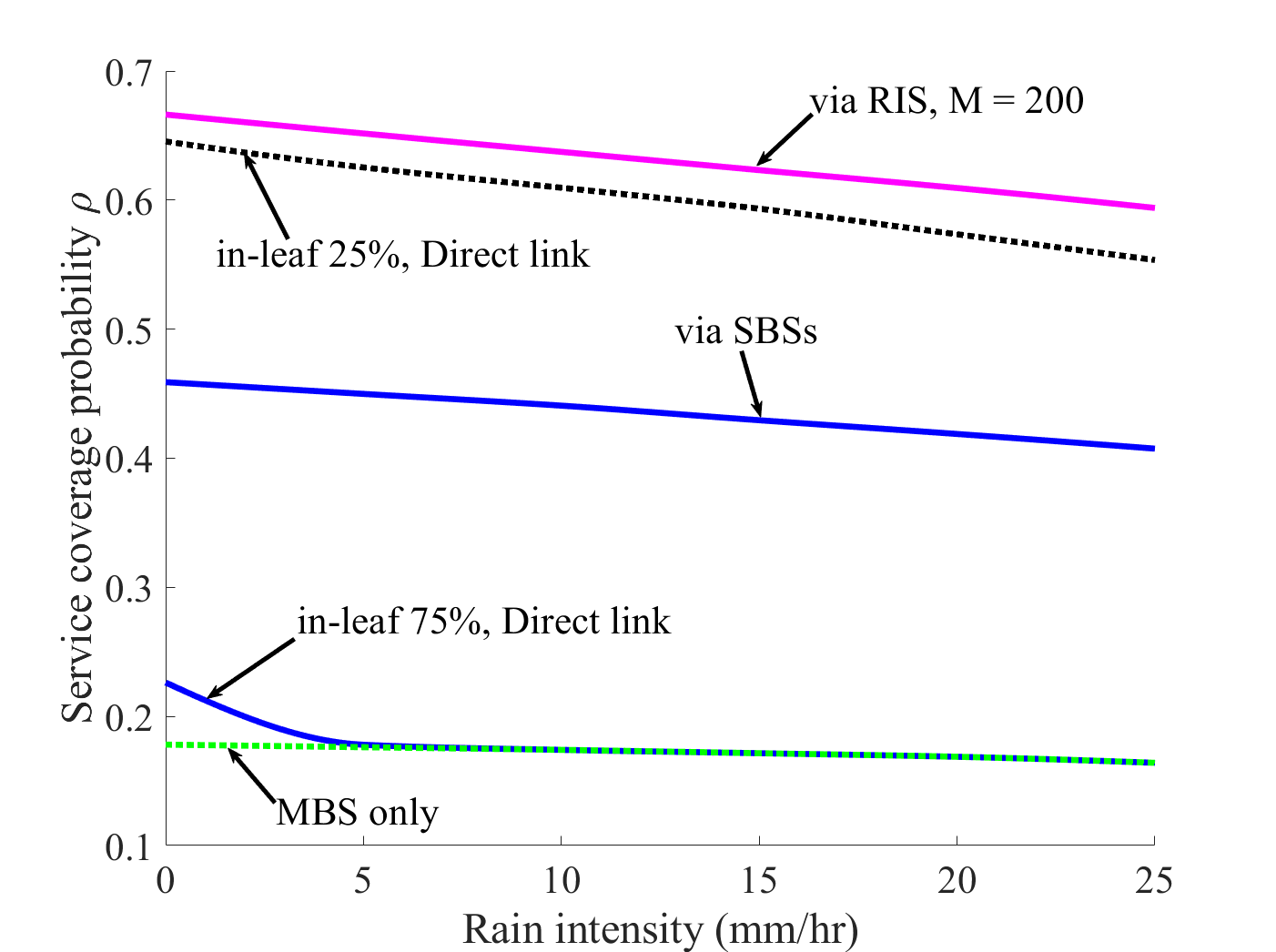}}
\caption{Network service coverage probability as a function of rain intensity in the presence of tree foliage.}
\label{stp_beamform2}
\end{figure}

\begin{figure}
\centerline{\includegraphics[width=3.4in]{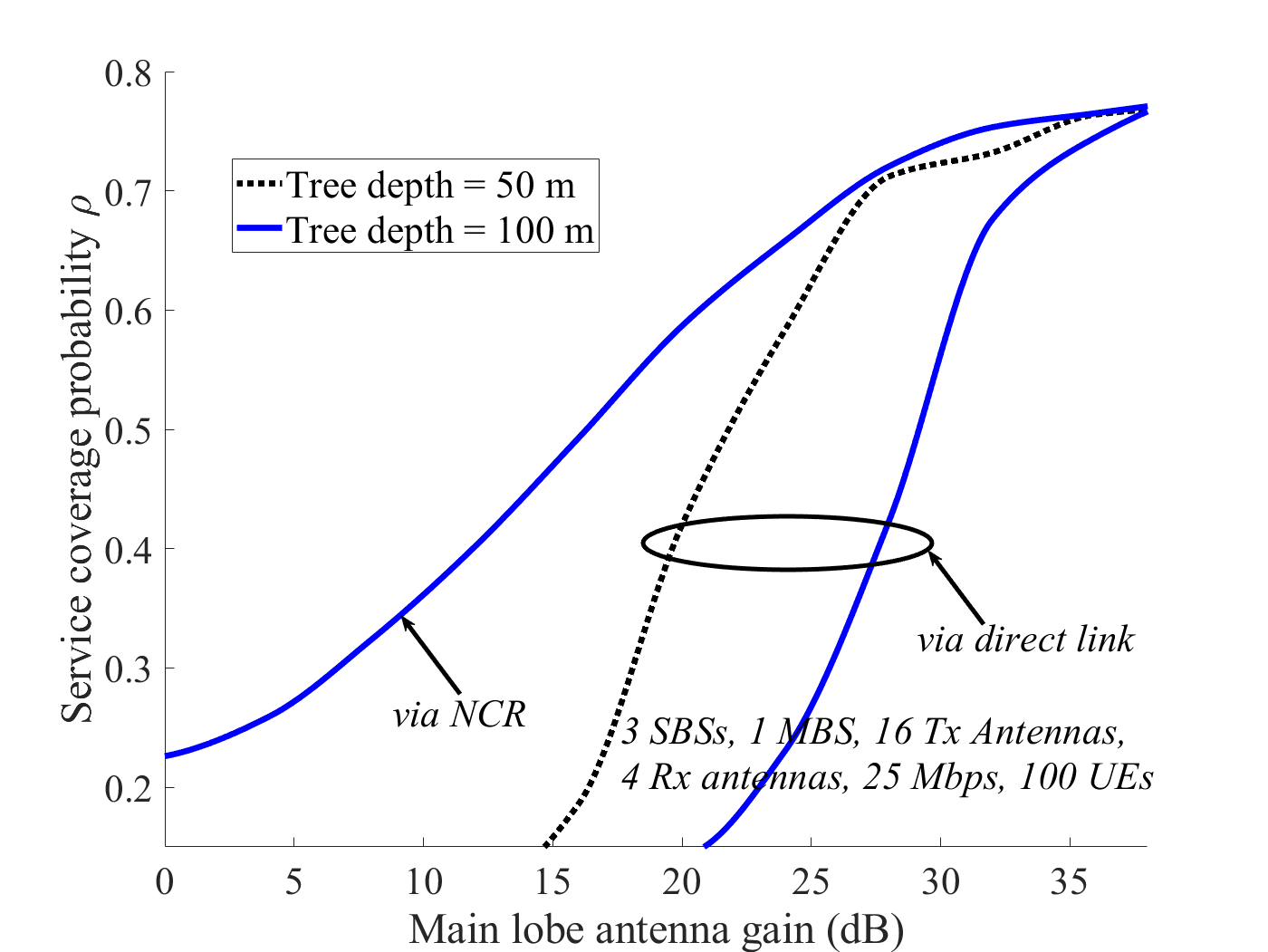}}
\caption{Network service coverage probability as a function of main lobe antenna gain.}
\label{gainfig}
\end{figure}

\subsubsection{Influence of Main Lobe Antenna Gain}
Figure \ref{gainfig} illustrates the effect of main lobe antenna gain on service coverage probability for two tree depths: 50 m and 100 m. The results reveal a significant improvement in service coverage probability with increasing antenna gain. For a tree depth of 50 m, service coverage probability reaches 0.7 at a main lobe gain of approximately 25 dB and saturates near 0.8 at higher gains. In contrast, for a tree depth of 100 m, service coverage probability shows a slower improvement, reaching only about 0.6 at a gain of 30 dB.

This disparity highlights the compounded effect of tree depth and antenna gain. Higher antenna gains effectively compensate for the path loss caused by tree obstructions, but their impact diminishes as tree depth increases. These findings emphasize the need for optimal antenna design and deployment strategies to enhance coverage in dense foliage environments.

\textcolor{black}{
Also, as seen in Fig. \ref{gainfig}, the use of NCR enhances service coverage probability compared to backhaul via direct links, particularly in low main lobe antenna gain scenarios. However, as the main lobe antenna gain increases, the direct link surpasses NCR, indicating that the effectiveness of NCR is higher in environments with lower antenna gain and higher tree depth. Here, the NCRs are placed at the same locations as of RISs and the maximum output power constraint,
amplification gain are set to 40 dBm and 100 dB, respectively. In particular, NCR and RIS have similar functionalities while there are detailed differences between them such as high power amplification and additional beamforming capabilities at the NCR (see \cite{carvalho2024network} for details about the NCR)}.

\subsubsection{Practical Implications}
Deploying RISs and SBS nodes effectively mitigates the adverse impacts of foliage and rain, by maintaining service coverage where direct links fail. RISs promise to provide cost-efficient backhaul solutions, while SBSs enhance reliability by distributing the network load. Also, we see the importance of selecting suitable frequencies based on deployment scenarios. Lower mmWave frequencies, such as 28 GHz, offer better coverage in dense environments, while higher frequencies, like 38 GHz, provide higher capacity but are more sensitive to obstructions. Optimizing antenna gains improve coverage, particularly in moderate foliage conditions. However, high obstructions require additional measures like RISs or SBSs for effective performance. Overall, we suggest that combining RISs, SBSs, selection of suitable carrier frequency, and advanced antenna designs ensures reliable connectivity especially in sub-urban wide-area IAB networks.

\section{Conclusion}

We studied the problem of RIS-assited IAB network to counter tree foliage loss over several seasons. This study provides valuable insights into the design of reliable and efficient IAB networks in challenging environments. The results demonstrate that deploying RISs/NCRs are useful for mitigating the effects of foliage, rain, and other environmental factors, ensuring robust service coverage, especially in wide-area IAB networks. The comparative analysis of 28 GHz and 38 GHz frequencies highlights the trade-offs between coverage and capacity, emphasizing the need for strategic frequency selection based on deployment scenarios. Additionally, optimizing the main lobe antenna gain further enhances coverage, particularly in moderate obstruction conditions. By integrating these strategies, one can achieve cost-effective and resilient IAB networks, catering to both urban and sub-urban deployment needs. Also, the proposed 
RIS-assisted scheme improves the service coverage probability throughout the year, regardless of the season and the carrier frequency used.

\section*{Acknowledgments}
This work was supported by the European Commission through the Horizon
Europe/JU SNS project Hexa-X-II (Grant Agreement no. 101095759).

\bibliographystyle{IEEEtran}
\bibliography{bibliography}

\end{document}